\documentstyle[twoside,fleqn,espcrc2,epsf]{article}

\newcommand{\eq}{\begin{equation}}
\newcommand{\en}{\end{equation}}
\newcommand{\eqa}{\begin{eqnarray}}
\newcommand{\ena}{\end{eqnarray}}

\newcommand{\AmS}{{\protect\the\textfont2
  A\kern-.1667em\lower.5ex\hbox{M}\kern-.125emS}}

\title{
\vspace{-1.8cm}
\hbox{}
{\small AUGUST 1996} \hfill {\small HLRZ 59/96}     \break
                                                              \break
Towards the glueball spectrum of full QCD\thanks{Presented by
    G.~Bali at Lattice '96, International Symposium on Lattice Field
    Theory, St.~Louis, USA, June 4 -- 8, 1996.}}

\author{SESAM collaboration: G.S.~Bali$^{\rm
        a}$, U.~Gl\"assner$^{\rm b}$,
        S.~G\"usken$^{\rm b}$, H.~Hoeber$^{\rm c}$, Th.~Lippert$^{\rm c}$, 
        G.~Ritzenh\"ofer$^{\rm c}$, K.~Schilling$^{\rm b,c}$, 
        G.~Siegert$^{\rm c}$ and A.~Spitz$^{\rm b}$\\[8pt]
{\rm $^a$}Department of Physics, The University, Highfield,
           Southampton SO17 1BJ, UK\\[8pt]       
{\rm $^b$}Physics Department, University of Wuppertal, D-42097
           Wuppertal, Germany\\[8pt]       
{\rm $^c$}HLRZ c/o Forschungszentrum J\"ulich, D-52425 J\"ulich,
          and DESY, D-22603 Hamburg, Germany}  
       
\begin{document}

\begin{abstract}
We present first results on masses of the scalar and tensor
glueballs as well as of the torelon from simulations of QCD with two
light flavours of Wilson fermions. The gauge configurations of extent
$16^3 32$ at $\beta = 5.6$ and $\kappa = 0.156, 0.157$ and 0.1575
have been generated as part of the SESAM collaboration programme.
The present lattice resolutions correspond to $a^{-1}=2.0$--2.3~GeV and
ratios $m_{\pi}/m_{\rho}\approx 0.83, 0.76$ and 0.71, respectively.
Studies on larger lattice volumes and closer to the chiral limit
are in progress.
\end{abstract}

\maketitle

\section{INTRODUCTION}
\noindent
Recently, increased attention has been paid on lattice glueball
computations~\cite{UKQCD,GF11}. It is demanding to include sea quarks
into such
studies as two new effects are expected to set in:
the scalar glueball becomes unstable and can decay into two
$\pi$'s (at sufficiently light sea quark mass). Also, mixing with
mesonic $I=0$ states might occur.
In this case, the glueball will just add
an additional excitation to the singlet state 
of the $L=1$, $J^{PC}=0^{++}$ meson quartet (nonet in full QCD with
three light flavours of sea quarks), which does
not fit into the constituent quark model.
In fact, one can equally well speak of a meson with valence quarks and
sea glue as of a
glueball with valence glue and sea quarks.
Only the couplings to certain decay channels can discriminate which of
these two features dominates in either state. Depending on the quark
masses, a glueball creation operator will project onto all these
states ($\pi\pi$ and mixed meson-glue states). One aim of the present
study is to look for unexpected or dramatic effects around
such thresholds.

In addition to glueballs, we study so called torelon states, i.e. flux
tubes wrapping around the lattice (with periodic spatial boundary
conditions). In case of the valence quark approximation, the mass of
such a state is (up to finite size corrections) expected to equal
$L_Sa\kappa$ where $L_S$ denotes the spatial extent of the lattice
(here: $L_S=16$) and $\kappa=K/a^2$ stands for the string
tension. When dynamical sea quark flavours are switched on, the
lightest such state ($A_{1g}$) can, at sufficiently large lattice
extent, decay into either a scalar meson or into two $\pi$'s or mix
with glueball states through intermediate mesonic states.
In the torelon we expect to be sensitive to ``string breaking''
effects.

It is well known from quenched studies that gluonic observables suffer
from large statistical fluctuations and small
signals~\cite{UKQCD,GF11,mtp}. Early studies with staggered sea
quarks~\cite{HEMCGC}
indicate that it might be extremely difficult to obtain significant
results on glueballs with dynamical fermions at all in realistic time. 
In view of the interesting physics involved, we wish to explore the
situation with our medium size sample of configurations.

\section{METHOD}
We compute matrices of zero-momentum
projected plaquette-plaquette and Wilsonline-Wilsonline correlators at
various fuzzing levels~\cite{mtp}. 
Two different fuzzing procedures have been applied for this purpose:
\begin{enumerate}
\item successive factor two blocking~\cite{mtp} with the straight
connection weighted by a coefficient $\alpha = 1$, 
in respect to the neighbouring four spatial
staples,
\item alternating between conventional APE smearing steps with
$\alpha=4$ (keeping the length of the fuzzed link constant)
and the above factor two blocking steps.
\end{enumerate}
Within procedure 1, the fuzzing levels are labeled by $n=1,\ldots,3$,
corresponding to links of effective length $2^n$ in lattice
units. Within procedure 2, two successive levels correspond to fuzzed
links of equal length, i.e.\ levels 0 and 1 yield links of extent 1
while 6 and 7 yield links of length 8.

Procedure 1 has been applied onto 2196 successive trajectories
at $\kappa = 0.157$, $V=16^332$, $\beta=5.6$
while procedure 2 has so far been applied to 204, 274 and 260
configurations, separated by about 10 trajectories, at $\kappa=0.156$,
$\kappa=0.157$ and $\kappa=0.1575$, respectively.
Prior to statistical analysis the data has been binned into
sufficiently large blocks to achieve stability of the error estimates
in respect to the block size, i.e.\ to avoid autocorrelations.
Statistical errors have been computed from the scatter between 500
bootstrap samples in either case.

Operators with quantum numbers $A_{1g}$ (the lightest torelon
state), $T_1^{++}$ (projecting onto the $J^{PC}=0^{++}$ glueball) and
$E^{++}$ ($2^{++}$) have been constructed at each fuzzing level.
After diagonalizing the $2\times 2$ submatrix
between the two basis states with best ground state overlap in each
case, masses of the glueball and torelon states are extracted.

Applying fuzzing procedure 1, levels 2 and 3 have been found
to project best onto the glueball ground states while just level 3
resulted in a satsifying overlaps for the torelon case.
For fuzzing 2, levels 6 and 5
(7 and 6 for the torelon) are optimal. After diagonalizing the
corresponding correlation
submatrices, we obtain ground state overlaps of $72\pm 7\%$ 
for the $0^{++}$ glueball and $70\pm 9\%$ for the torelon by use of
fuzzing 1. With procedure 2, glueball overlaps of $71\pm 6\%$,
$81\pm 5\%$ and $79\pm 5\%$ and torelon overlaps of
$93\pm 8 \%$, $98\pm 6\%$ and $89\pm 4\%$ have been achieved for the
three $\kappa$ values, respectively, 
resulting in an earlier onset of plateaus in local masses and, thus,
reduced statistical errors of final results.

\section{RESULTS}
\begin{figure}[htb]
\leavevmode
\epsfxsize=7.5cm\epsfbox{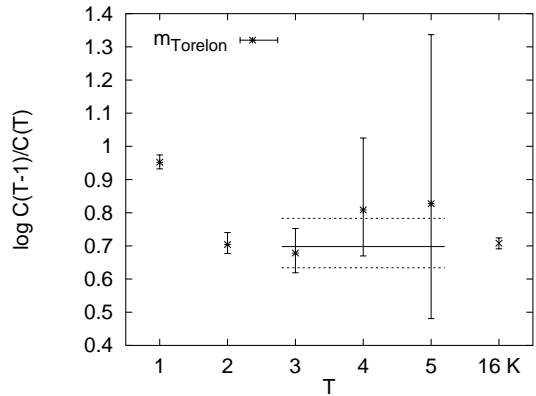}
\vskip -1cm
\caption{Torelon effective mass (fuzzing procedure 1). The rightmost
  point is the expectation from the static potential.}
\label{fig1}
\end{figure}

\begin{figure}[htb]
\leavevmode
\epsfxsize=7.5cm\epsfbox{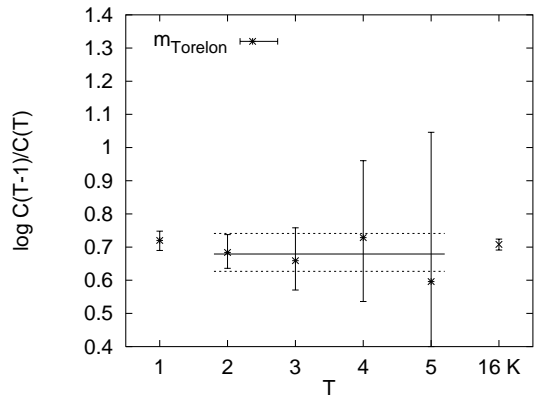}
\vskip -1cm
\caption{Torelon effective mass (fuzzing procedure 2).}
\label{fig2}
\end{figure}

In Figs.~\ref{fig1} and \ref{fig2} we
compare our results for the torelon mass at $\kappa=0.157$
from the two fuzzing procedures. The numerical values are
$M(A_{1g})=0.698^{+85}_{-65}$ and $0.679^{+62}_{-52}$,
in agreement with an expectation of $16 K = 0.712^{+16}_{-17}$, as
computed from the linear slope of the interquark potential~\cite{sesam}. Note,
that the leading order finite size effect, $\pi/(3L_S^2)\approx 0.004$,
is small.

\begin{figure}[htb]
\leavevmode
\epsfxsize=7.5cm\epsfbox{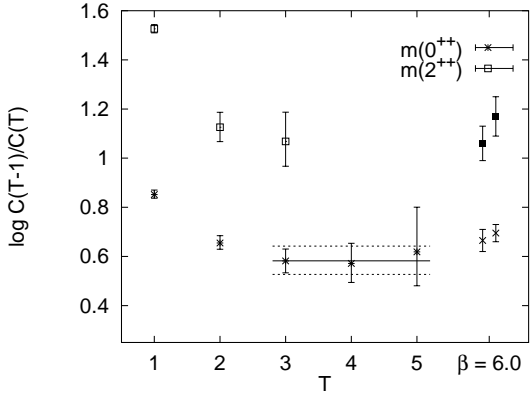}
\vskip -1cm
\caption{Effective glueball masses (fuzzing procedure 1). The
  rightmost data points are expectations from quenched simulations at
  $\beta=6.0$~[3,6].}
\label{fig3}
\end{figure}

\begin{figure}[htb]
\leavevmode
\epsfxsize=7.5cm\epsfbox{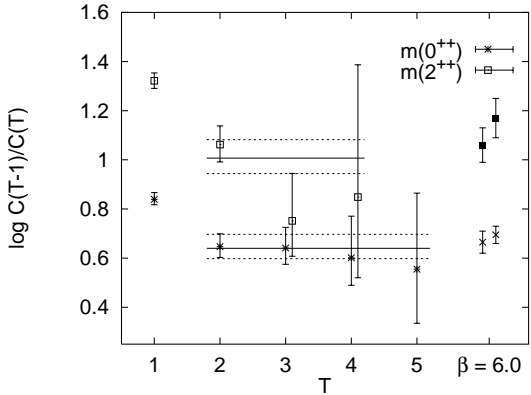}
\vskip -1cm
\caption{Effective glueball masses (fuzzing procedure 2).}
\label{fig4}
\end{figure}

In Figs.~\ref{fig3} and \ref{fig4} we relate our $\kappa=0.157$
results for glueball masses to those expected from quenched
simulations at $\beta=6.0$~\cite{mtp,UKQCD2}, i.e.\ at similar
lattice spacing ($a^{-1}\approx 2.1$~GeV).
For comparison, the quenched data points have been scaled by the ratio
of the corresponding $r_0$-values: $5.35/5.55$~\cite{sesam} within the
figures. Again, the two
fuzzing procedures yield consistent values.

\begin{table}[hbt]
\setlength{\tabcolsep}{0.8pc}
\newlength{\digitwidth} \settowidth{\digitwidth}{\rm 0}
\catcode`?=\active \def?{\kern\digitwidth}
\caption{Torelon masses.}
\label{tab1}
\begin{tabular}{lrrr}
\hline
$\kappa$&$m(A_{1g})a$&$L_Sa\kappa$&$2m_{\pi}a$\\
\hline
0.1560&0.80(7)&0.82(2)&0.890(2)\\
0.1570&0.68(8)&0.71(2)&0.677(2)\\
0.1575&0.58(4)&0.63(2)&0.576(9)\\
\hline
\end{tabular}
\end{table}

In Table~\ref{tab1}, we have collected results of the torelon mass
as extracted by use of fuzzing procedure 2 for all three $\kappa$
values. For comparison, we have included the expectation from the
(effective) slope of the interquark potential at large distance,
$\kappa$~\cite{sesam}, as well as the mass of two (noninteracting)
$\pi$'s~\cite{sesam2}. Note,
that at $\kappa=0.1575$ we are traversing the $\pi\pi$ threshold in
the sense that the $\pi$ turns lighter than half the
torelon mass expectation from the potential. Thus, it is not clear
whether our ground state still corresponds to the torelon or to a
two pion state. This might explain why the torelon seems to become
somewhat lighter than the expectation in this case.

\begin{table}[hbt]
\setlength{\tabcolsep}{1.4pc}
\catcode`?=\active \def?{\kern\digitwidth}
\caption{Glueball masses.}
\label{tab2}
\begin{tabular}{lrr}
\hline
$\kappa$&$m(0^{++})$a&$m(2^{++})a$\\
\hline
0.1560&0.70(6)&1.14(12)\\
0.1570&0.64(5)&1.01 (7)\\
0.1575&0.56(6)&1.00 (7)\\
\hline
\end{tabular}
\end{table}

\begin{table}[hbt]
\setlength{\tabcolsep}{0.8pc}
\catcode`?=\active \def?{\kern\digitwidth}
\caption{Comparison with quenched results.}
\label{tab3}
\begin{tabular}{lrr}
\hline
$\kappa$&$m(0^{++})r_0$&$m(2^{++})/m(0^{++})$\\
\hline
0.1560&3.60(35)&1.62(21)\\
0.1570&3.55(33)&1.58(17)\\
0.1575&3.28(38)&1.78(21)\\\hline
$\beta=6.0$&3.69(23)&1.67(15)\\
$\beta=\infty$&4.22(14)&1.40(15)\\
\hline
\end{tabular}
\end{table}
In Table~\ref{tab2} results for the scalar and tensor glueball masses
are displayed. Note, that the value $m(0^{++})a=0.56(6)$, 
obtained at $\kappa = 0.1575$, is consistent with
$2m_{\pi}a=0.58(1)$~\cite{sesam2},
opening this decay channel.
In Table~\ref{tab3}, we compare unquenched
with quenched results~\cite{mtp}
obtained at $\beta=6.0$ as well as quenched results from
Refs.~\cite{UKQCD,GF11,mtp,UKQCD2} that
have been extrapolated to the continuum ($\beta=\infty$)
quadratically in $a$.
The scale $r_0^{-1}$ from the interquark potential corresponds to
400~MeV. All results are in agreement with the quenched data at
$\beta=6.0$. Deviations from the continuum limit are expected to
depend linearly on $a$ with dynamical Wilson quarks.

\section{CONCLUSIONS}
We have studied masses of the scalar and tensor glueballs as well as
of the torelon with two flavours of Wilson quarks at $\beta=5.6$. All
results are consistent with those of quenched studies at similar lattice
spacings. The statistical errors turn out to be encouraging and will
compete with their quenched counterparts, once we have arrived at final
statistics. Between $\kappa=0.157$ and $\kappa=0.1575$, corresponding to
ratios $m_{\pi}/m_{\rho}=0.76$ and 0.71~\cite{sesam2}, the $\pi$
becomes lighter than half the torelon mass and string breaking is
expected to set in on $L_S=16$ lattices. Around
$\kappa\approx 0.1575$ the scalar glueball becomes unstable, which
means that this region deserves further attention.
Computations on $24^340$ lattices at $\kappa=0.1575$ and
$\kappa=0.158$ are in progress.

\section*{ACKNOWLEDGEMENTS}
The present work has been supported by DFG grants Schi 257/1-4 and Schi
257/3-2, EC project SC1*-CT91-0642 and EC contract CHRX-CT92-0051.
GB acknowledges support by EU grant ERB~CHBG~CT94-0665.

\end{document}